\documentclass[fleqn,twoside]{article}

\usepackage{graphicx}
\usepackage[figuresright]{rotating}
\usepackage{amsmath}
\usepackage{amssymb}
\usepackage{espcrc2}
\usepackage{mcite}

\title{Instantons and chiral symmetry breaking in SU($N$) gauge theories.}

\author{N. Cundy\address[TP]{Theoretical Physics, 1 Keble Road,
        Oxford, OX1 3NP, United
        Kingdom.}\thanks{Speaker}
\thanks{NC is supported by PPARC
        grant PPA/S/S/1999/02872}, %
        M. Teper\addressmark,  %
        and U. Wenger\addressmark
\thanks{UW acknowleges financial support from
        PPARC SPG}
}
       
\begin{document}

\begin{abstract}
We address the question of whether the low modes of the Dirac operator
are caused by topological objects such as instantons in SU($N$) gauge
theories. We study the pseudo-scalar density of these modes, finding the size distributions of the instantons,
and comparing it with the underlying gauge field. We find that,
although the near-zero modes of the Dirac operator depend on topology
for all $N$, their small instanton content decreases as $N$ increases.
\vspace{-0.5cm}
\end{abstract}

\maketitle

\section{INTRODUCTION}
The mechanism behind spontaneous chiral symmetry breaking (S$\chi$SB) is an
unsolved mystery in QCD. A popular hypothesis is that a weakly
interacting dilute gas of instantons and anti-instantons could create
a number of small non-zero eigenvalues of the Dirac
matrix~\cite{DiaPet1}.  By the Banks-Casher
relation~\cite{Banks-Casher}, this leads to a non-zero chiral
condensate, and therefore  S$\chi$SB. Using chiral Dirac
operators, such as the overlap operator~\cite{Neuberger1}, one can
examine the scalar and pseudo-scalar densities of the low-lying
eigenmodes of the Dirac operator numerically, searching for
instantons. Recent studies of these modes in SU(3) have found a lumpy
self-dual
structure~\cite{Horvath1,Horvath2,DeGrand,Blum,Hip,Gattringer}. Although this is predicted by the instanton liquid model (ILM),
there is continuing discussion about whether these lumps are
instantons.

It has been suggested that ILMs are inconsistent with large $N$
expansions. At large $N$, it has been proposed that the instanton
weight in the partition function vanishes exponentially~\cite{Witten}
(but see ~\cite{Teper,Schafer}), while effects from large quantum
fluctuations decay according to a power law, probably $1/N$. This
implies that instanton effects are insignificant compared to quantum
effects at large $N$, and it is unlikely that S$\chi$SB in SU($N$)
gauge theories for large $N$ is caused by instantons, and unless there
is a phase transition, S$\chi$SB in SU($3$) will not be caused by
instantons either. Partly because of this, some people believe that
the instanton liquid is destroyed by quantum effects~\cite{Horvath1,Horvath2}.

We are investigating what happens to the lumpy structure in the low
modes of the Dirac operator when we change the number of
colours~\cite{cundy}. To do so, we generated configurations on a $12^4$
lattice, for SU($N$) gauge theories ($N= 2,3,4,5$), keeping the lattice
spacing fixed at $a=0.12$ fm by keeping a constant string
tension~\cite{Lucini}.
\begin{figure*}
\begin{tabular}{l l}
\includegraphics[width = 7cm,height =
4.9cm]{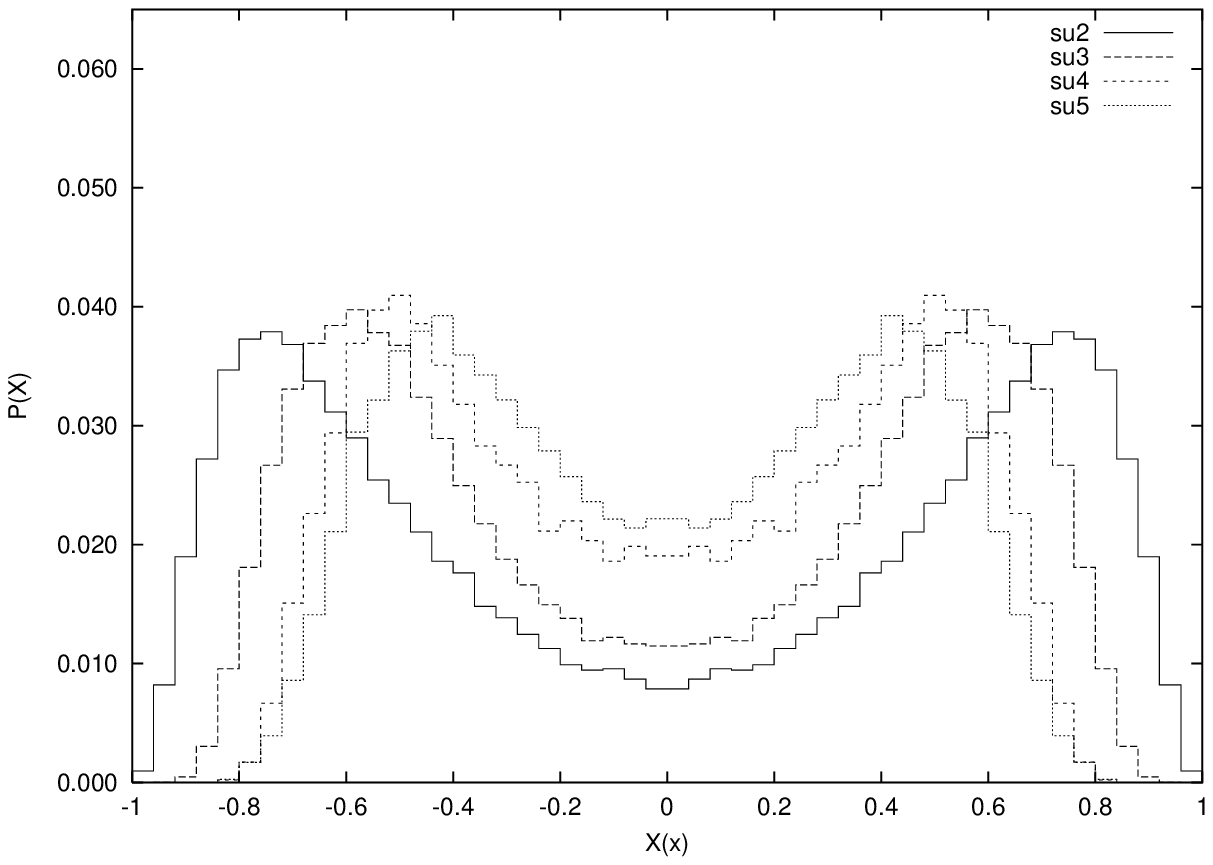}
\includegraphics[width = 7cm,height =
4.9cm]{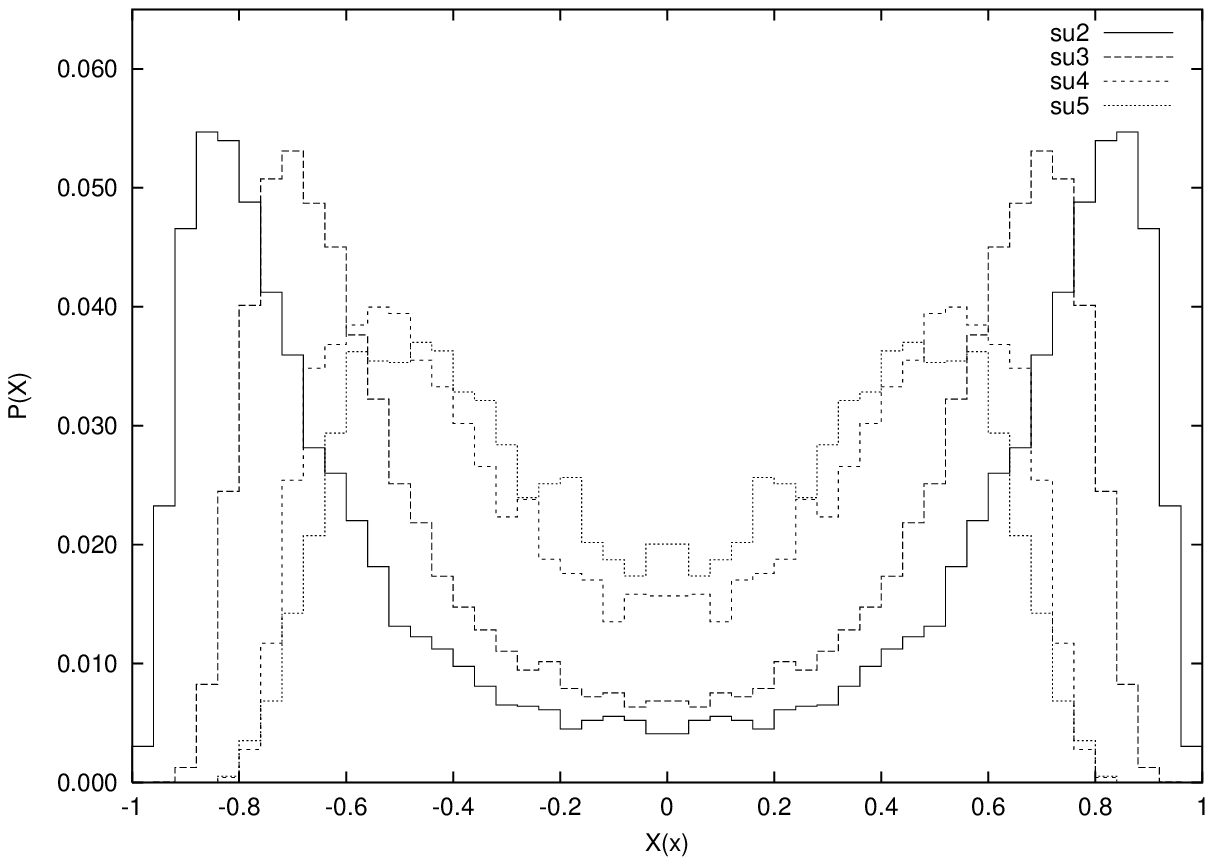}&
\end{tabular}
\vspace{-1cm}
\caption{Local chirality distributions for $\lambda^2 < 0.1$ (left) or
$\lambda^2 < 0.03$ (right). 
}\label{fig4}
\vspace{-0.5cm}
\end{figure*}

\section{TOPOLOGICAL CHARGE AND SMALL INSTANTONS}
There are two different measures of topological charge which we used.
The fermionic charge, $Q_f$, is defined as the difference between the number of
negative and positive chirality zero-modes of the Dirac
operator~\cite{Atsing}. Secondly, we can use a lattice equivalent of
the continuum topological charge density, $F\tilde{F}(x)$, as a 
(``field theoretic'') topological charge density, $q(x)$. We have to
cool the gauge field before calculating $q(x)$ to suppress UV
fluctuations. A gauge charge, $Q_g$, is the sum of $q(x)$ over all
lattice sites. In the continuum, $Q_g$ and $Q_f$ are equal. However,
lattice artifacts can lead to $Q_g\neq Q_f$. We found considerably
more discrepancies in SU(2) and SU(3) than in SU(4) and SU(5). In each
of the configurations where there was a discrepancy, there was also a
small instanton of size $\rho <0.3$ fm (found by looking at the size of
the peaks in $q(x)$)~\cite{cundy}. We concluded that for the lattice
spacings we used, the overlap operator cannot resolve instantons below
this size.
\section{SIZE DISTRIBUTIONS}
For each mode, we identified the peaks in the pseudo-scalar density.
Using classical instanton formulae for the height and shape of the peaks~\cite{cundy}, we calculated size
distributions for the instantons contributing to the eigenmodes.  We saw
that as $N$ increases, the typical instanton size becomes slightly
larger, and small instantons ($\rho \lesssim 0.4$ fm) disappear, in
agreement with theoretical predictions~\cite{Schafer}.

\section{LOCAL CHIRALITY}\label{sec:lcop}
The local chiral orientation angle, $X$~\cite{Horvath1}, is defined at each lattice site as
{\small
\begin{gather}
X = \frac{4}{\pi}\left(\arctan\left(\left[\frac{\psi_{-}\psi^{\dagger}_-}{\psi_{+}\psi^{\dagger}_+}\right]^{\frac{1}{2}}\right)-1\right).
\end{gather}}
$\psi_{+}$ and $\psi_{-}$ are the projections of the eigenmodes of the
Dirac matrix onto the $\pm$ chiral sectors. An isolated
(anti-)instanton, will have $X = \pm 1$, and
for a classical ILM, this would dominate the local chirality distribution, a histogram of
the values of $X$ on the lattice sites with the highest scalar density. If the
modes are not caused by topology, then the distribution would be centred
around $X = 0$. Overlapping instantons, instantons in a background of
quantum fluctuations, or a background of large quantum fluctuations
will be between the two extremes. Figure \ref{fig4} shows a histogram of the values of $X$ on the 2\% of
lattice sites with the highest scalar density, averaged over two
different ranges of eigenvalues $\lambda$.

As $N$ increases, or the eigenvalue increases, the peaks in the local
chirality distributions move towards zero, behaving less like the
expected distribution from classical instantons. If the shape of the
distributions is determined by small instantons, then excluding
those SU(3) configurations which contain narrow instantons would make
the distribution behave more like SU(4), since there are fewer
small instantons there. However we did not
observe a significant effect when we tried this, and so we conclude that small instantons
are not responsible for the different shape of the chirality
distributions for the various gauge groups.

\section{AUTO-CORRELATORS}\label{sec:ac}
\begin{figure*}
\begin{tabular}{c c}
\includegraphics[width =
7cm,height = 4.9cm]{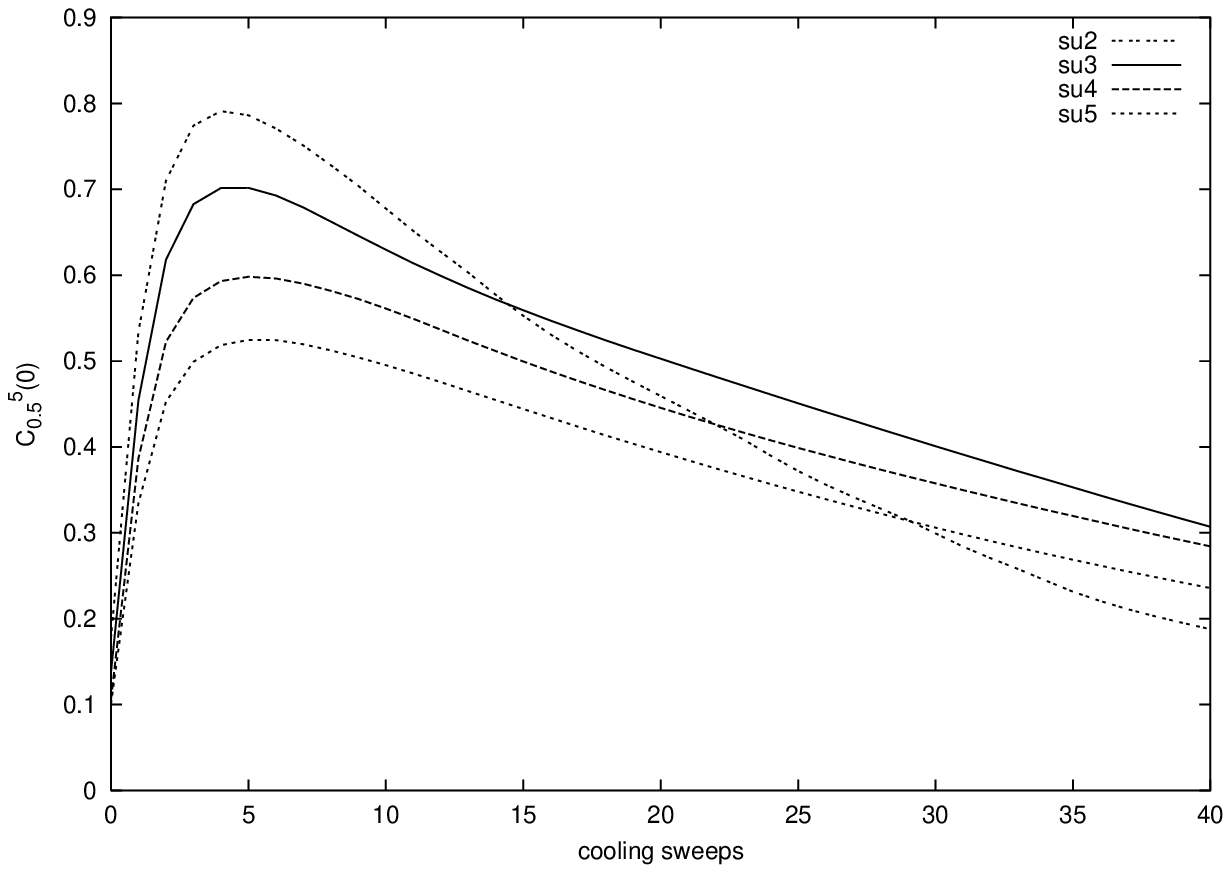}&
\includegraphics[width =
7cm,height = 4.9cm]{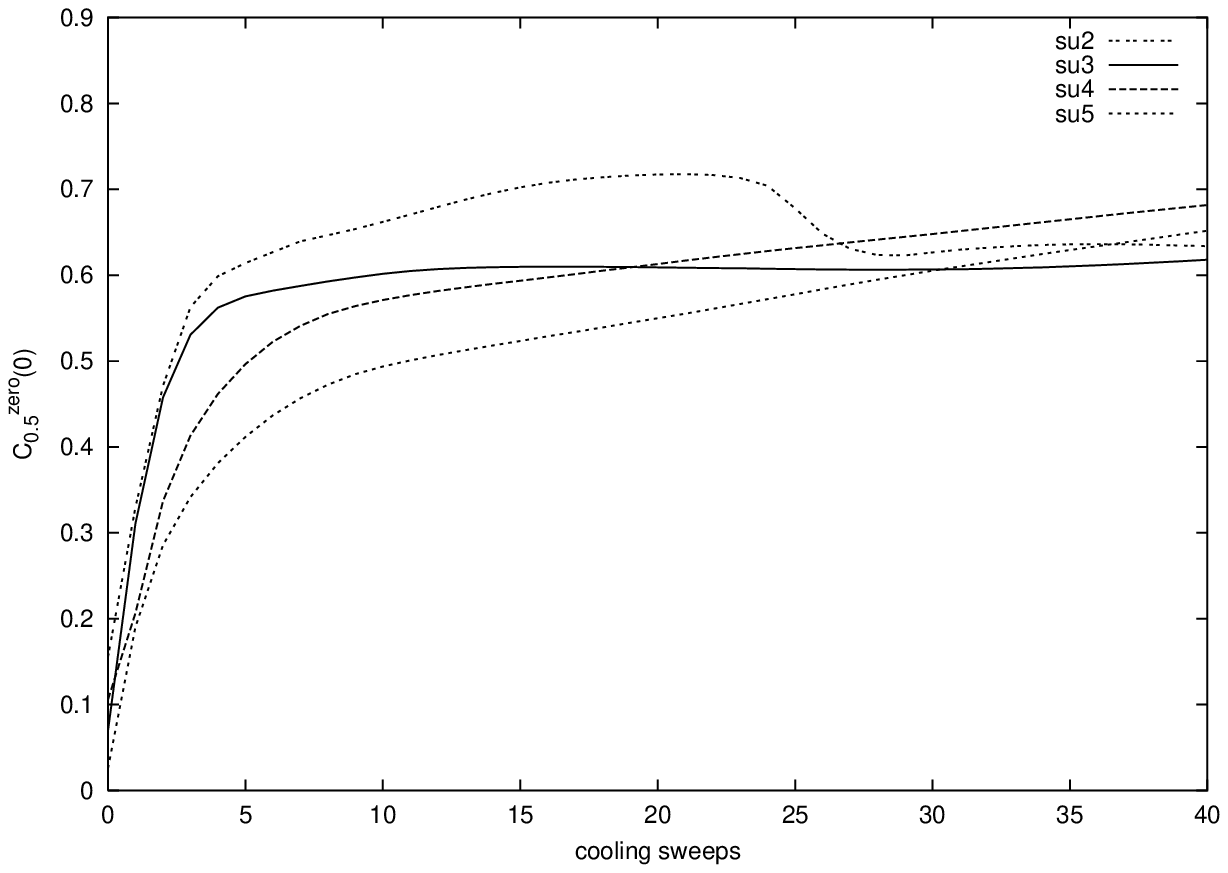}\\
\end{tabular}
\vspace{-1cm}
\caption{$C_{0.5}^5(0)$ for non-zero modes (left) and zero-modes with
$|Q_g|=1$ (right). The plots are ensemble averages, and are shown as a
function of cooling sweeps at which the gauge charge was
calculated. In one of the SU(2) $|Q_g| = 1$ configurations, an
instanton is destroyed by cooling at 25 sweeps.  }\label{fig6}
\vspace{-0.5cm}
\end{figure*}

To see whether a mode is influenced by topology, one can compare it
with the underlying gauge field. We used a scale-invariant
(dimensionless) auto-correlator which is (in the continuum)
independent of the instanton size, {\small\begin{align} C_d^5(0)\equiv
\int d^4 x &\left|\psi^{\dagger}(x)\gamma^5\psi(x)\right|^d
\text{sign}\left(\psi^{\dagger}(x)\gamma^5\psi(x)\right)\nonumber\\
&\left|{q(x)}\right|^{1-d}\text{sign}\left(q(x)\right)\label{eq:sdq:sdqdef1},
\end{align}}
where $d$ lies in the range $0\le d\le1$. If the pseudo-scalar density
is correlated with topological charge (for example, if the eigenmode
contains an instanton/anti-instanton pair), then $C_d^5(0)>0$. If the
eigenmode is not influenced by topology, then there will be no
correlation between the pseudo-scalar and topological charge densities,
giving $C_d^5(0)\sim 0$. We plot the auto-correlator as a function of
the number of cooling sweeps at which we calculate $q(x)$ (figure
\ref{fig6}). The auto-correlator quickly rises to a maximum, and then decreases as the cooling distorts the
gauge field. We can calculate, by modelling artificial instanton
configurations, that for a classical instanton/anti-instanton pair,
the auto-correlator should give a value $\sim 1.1$. Overlapping
instantons, instantons in a background of quantum fluctuations, or
non-instanton topological fluctuations would give a smaller value than
this. The measured
value, at 6 cooling sweeps (the maximum for the non-zero modes), is
smaller than the expected classical value, and decreases as we increase the number
of colours. Instantons seem to contribute less to the modes as we
increase $N$. On the other hand, the zero modes, which we know are
caused by topology, show a similar decrease. The ratio between the
auto-correlators (at 6 cooling sweeps) for the non-zero modes and for the zero-modes is
constant ($\sim 1.4$). This suggests that topology is important in
chiral symmetry breaking $\forall$ SU($N$), but that classical
instantons become less important as we increase $N$.

\section{CONCLUSIONS}
We found that discrepancies between the fermionic and gauge charges
are caused by small instantons ($\rho < 0.3$ fm in
this calculation). As we increase $N$, small instantons ($\rho
\lesssim 0.4$ fm) are suppressed. Examination of the auto-correlator
suggests that chiral symmetry breaking at large $N$ does have a
topological origin. However, as we increase $N$, semi-classical
instantons seem to become less important. Our results, from sections
\ref{sec:lcop} and \ref{sec:ac}, imply that if the lumps in the low
modes of the Dirac operator are instantons, then as we increase $N$, the instantons either overlap more heavily, and/or
quantum fluctuations become more important.

\bibliographystyle{elsart-num} 
\bibliography{lattice_2002_pro}

\end{document}